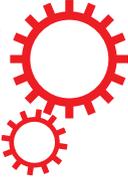

www.nature.com/scientificreports

# OPEN

# All-optical central-frequency-programmable and bandwidth-tailorable radar

Weiwen Zou[1,2], Hao Zhang[1], Xin Long[1], Siteng Zhang[1], Yuanjun Cui[1] & Jianping Chen[1,2]

Radar has been widely used for military, security, and rescue purposes, and modern radar should be reconfigurable at multi-bands and have programmable central frequencies and considerable bandwidth agility. Microwave photonics or photonics-assisted radio-frequency technology is a unique solution to providing such capabilities. Here, we demonstrate an all-optical central-frequency-programmable and bandwidth-tailorable radar architecture that provides a coherent system and utilizes one mode-locked laser for both signal generation and reception. Heterodyning of two individually filtered optical pulses that are pre-chirped via wavelength-to-time mapping generates a wideband linearly chirped radar signal. The working bands can be flexibly tailored with the desired bandwidth at a user-preferred carrier frequency. Radar echoes are first modulated onto the pre-chirped optical pulse, which is also used for signal generation, and then stretched in time or compressed in frequency several fold based on the time-stretch principle. Thus, digitization is facilitated without loss of detection ability. We believe that our results demonstrate an innovative radar architecture with an ultra-high-range resolution.

Like other wireless equipment, a conventional microwave and millimetre-wave radar system operates efficiently only on a pre-designed band and inevitably requires up-conversion and down-conversion procedures[1–4]. Radar waveforms at the baseband are synthesized by digital electronics and then up-converted to the carrier frequency. During reception, the radar echoes must be electronically down-converted before signal processing. Metamaterials with distinctive electromagnetic scattering characteristics[5–7] have been proposed as a method of evading radar detection, with the target covered with a metamaterial cloak and electromagnetic waves absorbed or scattered by the cloak. Hence, the radar cross-section is effectively reduced, and the target cannot be perceived. However, this type of cloak is hardly perfect because it can work only at a particular frequency band. To improve performance in detecting a stealth target, modern radar should be reconfigurable and capable of working at multi-bands. Because the frequency characteristics of the microwave components are usually fixed, a radar system capable of working at multi-bands should have an array of generating/transmitting and receiving/processing units. Because such systems are composed of several sets of frequency conversion components, the systems are complicated. A linearly chirped waveform is one of the most commonly used radar signals and extensively employed in modern radar systems to improve the range resolution, dynamic range, and/or signal-to-noise ratio based on pulse compression technology[1–2]. Photonics-assisted radio-frequency (RF) technology or microwave photonics[8–9] can improve the flexibility of signal generation for a linearly chirped waveform[10–22] to overcome the bandwidth limits and time jitter of modern electronic or microwave devices. Therefore, microwave photonics has been considered a key technology for future radars[23–24]. In a pioneering demonstration[23], a photonics-based receiver and transceiver have been incorporated into a coherent radar architecture, and the capabilities have been verified in a real detection test. Using this system, direct RF signal generation and reception are simultaneously realized based on a single mode-locked laser (MLL), thus avoiding electrical frequency conversions. The digitization fidelity of the photonic analogue-to-digital conversion (ADC) and agility of the carrier frequency increase the competitiveness of the architecture relative to its all-electronic counterpart. It is worth mentioning that the coherent radar still requires an electro-optic modulation of the baseband signal, and a compromise can be found between the achievable bandwidth and the unambiguous detection distance when increasing the repetition rate of the MLL.

[1]State Key Laboratory of Advanced Optical Communication Systems and Networks, Department of Electronic Engineering, Shanghai Jiao Tong University, Shanghai 200240, China. [2]Shanghai Key Lab of Navigation and Location Services, Shanghai Jiao Tong University, Shanghai 200240, China. Correspondence and requests for materials should be addressed to W.Z. (email: wzou@sjtu.edu.cn) or J.C. (email: jpchen62@sjtu.edu.cn)

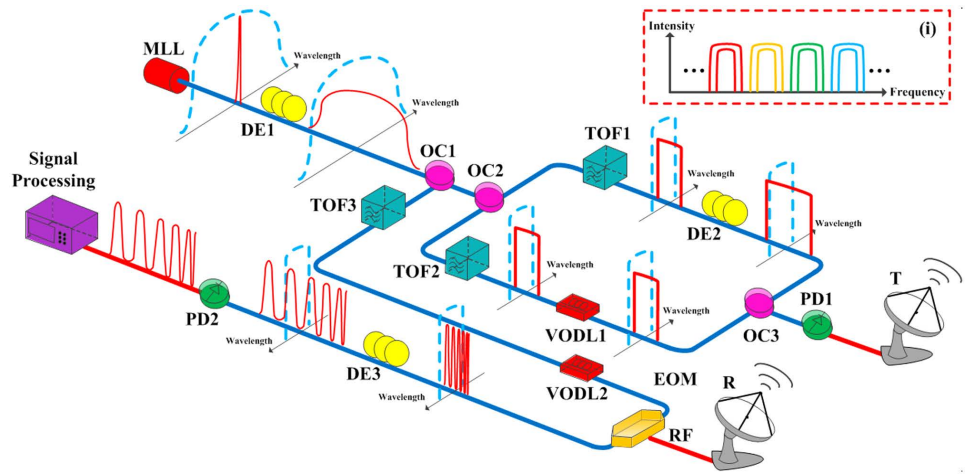

**Figure 1. Architecture of the all-optical central-frequency-programmable and bandwidth-tailorable radar.** An MLL with broadband optical spectra is used as the optical source for both signal generation and reception. DE1 maps the broad optical spectra of the short pulse train to the time domain. After it is coupled into two arms and filtered by two TOFs, the signal in each arm becomes an approximately rectangular profile. The signal in the first arm is temporally stretched wider than that in the second arm because of the addition of DE2. The detected linearly chirped signal is emitted by the transmitting (**T**) antenna. The echo pulses received by the receiving (**R**) antenna are directed to the RF port of the EOM. TOF3 filters the pre-chirped optical pulse train from OC1, and VODL2 compensates for the time difference between the optical and electrical signals into the EOM. DE3 further stretches the modulated optical signal. After the opto-electronic conversion, the received echoes are temporally stretched and compressed in frequency. The signal processing module digitalizes the time-stretched signal. The inset (i) denotes how different carrier frequencies and/or different sweep bandwidths are generated.

In this article, we propose an all-optical central-frequency-programmable and bandwidth-tailorable radar architecture. Linearly chirped signals are optically generated or received without the assistance of electronic up-conversion and down-conversion procedures. By simply adjusting the optical devices, the proposed architecture can rapidly adjust among different operating bands. Such frequency agility is appealing in diverse applications. For instance, the current cloaking technique cannot ensure full bandwidth coverage; thus, the radar cross-section might be large at certain bands, which greatly increases the possibility of successfully detecting the cloaked target when using this architecture. During reception, the received radar echoes are optically stretched in the time domain and compressed in the frequency domain. Hence, electronic down-conversion is avoided, and the ADC sampling rate is essentially reduced.

## Results

**Principle and experimental setup.** Figure 1 shows the setup and principle of the proposed all-optical central-frequency-programmable and bandwidth-tailorable radar. Only one MLL is used as the optical source for both signal generation and reception, thus ensuring strict coherence within the entire system. For the signal-generation component, one dispersion element (DE1) with a relatively large dispersion value is laid after the MLL. Because of the wide optical spectrum of the MLL, the short optical pulse train is broadened or chirped in the time domain at the same repetition rate. The time profile is the same as the optical spectrum because of the linear dispersion used. An optical coupler (OC1) divides the chirped optical signal into two parts for signal generation and reception, and another optical coupler (OC2) divides the optical pulse train into two arms for signal generation. Two tunable optical filters (TOFs) with approximately rectangular filtering characteristics are placed in two arms to tailor certain segments of the optical spectra. According to the Fourier transformation or wavelength-to-time mapping introduced by dispersion, the filtered wideband optical spectra are linearly mapped to the time domain. Therefore, the temporal and spectral profiles of the filtered pulses in the two arms are almost rectangular. A second dispersion element (DE2) with a small dispersion value is added to the first arm to induce a dispersion difference between the two arms so that the temporal duration of the pulse in the first arm becomes wider (or narrower) than that in the second arm according to the dispersion sign of DE2. To compensate for the time difference between pulses in the two arms, a variable optical delay line (VODL1) is inserted in the second arm. After heterodyning, or the beating of two differently dispersed optical pulses at the optical coupler (OC3) and opto-electronic conversion at a high-speed photo-detector (PD1), the generated RF signal is linearly chirped in frequency. The carrier frequency and sweep bandwidth are determined by differences between the central wavelengths and filtering bandwidths of the TOFs, respectively. This approach, which does not require any electronic devices, can be used to easily tailor different carrier frequencies and sweep bandwidths by simply tuning the two TOFs, which is schematically shown in the inset (i) of Fig. 1.

In the reception component, the frequency of the echoes is tailored and the bandwidth is compressed based on the time-stretch principle. The time-stretch principle has been successfully used in photonic ADC[25–28]. As shown in Fig. 1, the received radar echoes are directed to the RF port of an electro-optic modulator (EOM), and they are modulated onto the pre-chirped and filtered optical carrier divided from the OC1. TOF3 with an approximately rectangular profile is added to filter the relatively flat wavelength range of MLL considering that flatter optical spectra can contribute to better amplitude uniformity. The filtering bandwidth of TOF3 is set much wider than that of TOF1 and TOF2 because the temporal duration of the radar echoes for multi-target detection is usually longer. A wider filtering bandwidth ensures that the echoes can be modulated onto the optical carrier. Another variable optical delay line (VODL2) is inserted between the TOF3 and EOM to compensate for the time difference between the optical carrier and the radar echoes to be modulated. A third dispersion element (DE3) with a larger dispersion value than DE1 is placed after the EOM to further chirp the modulated optical pulses. When detected by a lower-speed photo-detector (PD2), the signals are time stretched and frequency compressed several fold compared with the original radar echoes. Therefore, the ADC sampling rate after the reception component can be lowered correspondingly.

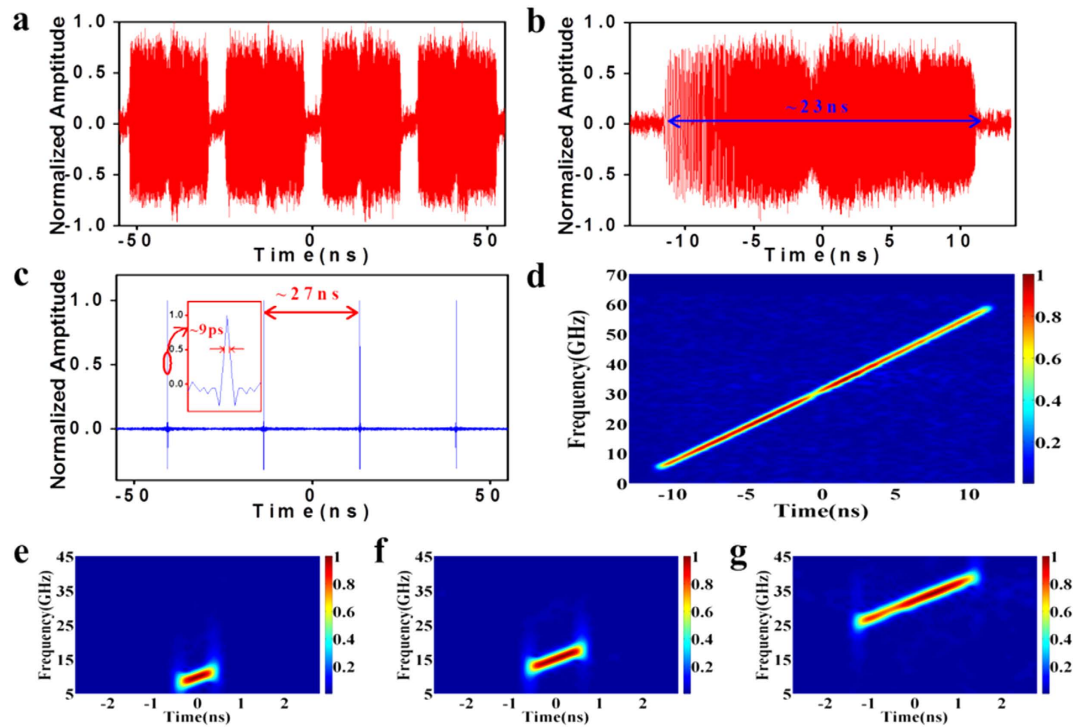

**Figure 2. Measurement results for the generated signals.** (**a**). Generated linearly chirped pulse train covering an ultra-broad wideband from 5 GHz to 60 GHz with a repetition rate of 37 MHz and a large TBWP of ~1265. (**b**). Magnified temporal waveform of the generated signal in (**a**,**c**). Autocorrelation of the generated linearly chirped pulse train in (**a**). The temporal width of the compressed pulse shows the pulse compression capability of the generated signal when used for detection. (**d**). Sshort-time Fourier transform (STFT) analysis of the generated signal in (**a**,**e**). STFT analysis of the generated X-band signal. (**f**). STFT analysis of the generated Ku-band signal. (**g**). STFT analysis of the generated Ka-band signal.

**All-optical radar signal generation.** As proof of concept, a wideband linearly chirped signal with a considerable time-bandwidth product (TBWP) and signals at three tailored frequency bands are demonstrated. Figure 2 shows the measurement results of the generated signals. Because we were constrained by the current testing condition, we have demonstrated a signal only under the bandwidth limit (~63 GHz) of the oscilloscope. Figure 2a–d demonstrates the results when the waveform is designed to cover a wide bandwidth with a repetition rate of 37 MHz (i.e., 27-ns period) as the MLL. The duration of each pulse lasts ~23 ns, and the short-time Fourier transform (STFT) analysis shows that the generated signal covers a bandwidth from 5 GHz to 60 GHz, thus resulting in a large TBWP of ~1265. This TBWP is the largest on record to our knowledge for methods based on the wavelength-to-time mapping principle. We also calculated the autocorrelation of each pulse to evaluate the pulse compression capability of the generated signal. Owing to the wide bandwidth, the pulse duration is compressed to ~9 ps with a period of ~27 ns. The carrier frequency and bandwidth are independently programmable by adjusting the central wavelength and filtering bandwidth of the two TOFs. Figure 2e–g shows the STFT analysis of the tailored signals in the X-band, Ku-band, and Ka-band. The waveforms in the three bands maintain the same chirp rate because they are, in principle, determined by the dispersion values of DE1 and DE2. In the demonstrated results, the 37 MHz pulse envelope has been eliminated; the detailed processing method can be found in the Methods.

**All-optical radar signal reception.** A single-target detection experiment was conducted to verify the proposed bandwidth-tailorable reception. As illustrated in Fig. 3a, the frequency components are compressed, and the temporal durations are expanded in the signal reception component. If DE1 and DE3 are the same type of optical fibres and $L_1$ and $L_3$ are their individual lengths, then the frequency components are tailored or compressed by a factor of M (=$L_3/L_1 + 1$), which is called the time-stretch factor[25], because the temporal duration of the echoes is expanded by M times with respect to the original duration $t_0$. If M is large enough, then the high-frequency signal can be converted to a low-frequency signal. The X-band linearly chirped signal described in Fig. 2e is employed as an example. Figure 3b–c show the schematic diagram and experimental layout, respectively. The all-optically generated X-band signal is amplified by a power amplifier (PA) and transmitted by a horn antenna. The received echo collected by another horn antenna is first amplified by a low noise amplifier (LNA) and goes through the signal reception component. The two horn antennas are positioned approximately side by side, and a metal plate is placed in front of the antennas as the reflective target. The filtering bandwidth of TOF3 is set much wider than that of TOF1 and TOF2, which leads to a longer temporal duration of the pre-chirped optical pulse and increases the possibility of successful modulation. Moreover, based on the relevance between the distance and arrival time of the radar echoes, VODL2 is adjusted to ensure the successful modulation of the radar echoes onto the pre-chirped optical carrier. Because the maximum range of VODL2 is limited, the short length of the optical fibre is correspondingly cascaded. A comparison of signals before transmission and after the time-stretched reception is provided in Fig. 3d–g. The waveform in Fig. 3d lasts ~1 ns, whereas that in Fig. 3e lasts ~5 ns, indicating a value of M = ~5. Figure 3f–g presents the STFT analysis results. Compared with the original transmitted signal, the time-stretched echo has been compressed ~5-fold. The central frequency of the signal after reception is located at ~2 GHz, and the linearly chirped feature has been retained. This demonstration shows the ability of the signal reception component to equivalently down-convert echoes and alleviate the sampling rate demands.

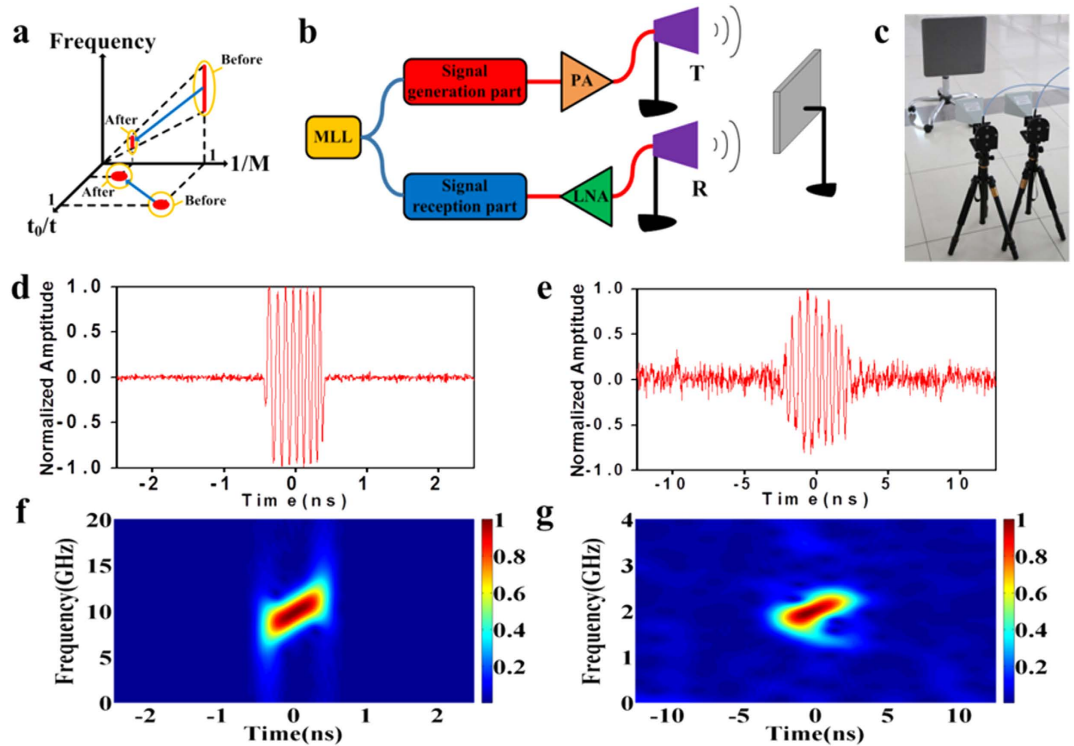

**Figure 3. Principle and measurement results for the time-stretched reception.** (**a**). Principle of time-stretched reception. (**b**). Schematic diagram of single-target detection. (**c**). Experimental layout of single-target detection. (**d**). Temporal waveform of the transmitted X-band signal. (**e**). Temporal waveform of the received X-band echo after the time-stretched process; the stretch ratio is M = ~5. (**f**). STFT analysis of the transmitted X-band signal. (**g**). STFT analysis of the received X-band signal after the time-stretched process; the stretch ratio is M = ~5.

**Radar detection ability.** The range resolution is usually used to characterize the detection ability. To better distinguish among two or more neighbouring targets, a linearly chirped waveform with a broader bandwidth is desired. The generated X-band signal is used again as an example for performing dual-target detection. The schematic diagram and experimental layout are illustrated in Fig. 4a–b. Two metal plates are placed side by side but separated by a distance of $\Delta D$ along the ranging direction. Figure 4c–d shows the normalized echoes after the compressed reception when $\Delta D$ is changed. The horizontal axis is the round-trip time for travelling. The reflected echoes are overlapped in the time domain in Fig. 4c when the targets are very close, which might result in ambiguity. When $\Delta D$ increases, the two reflected echoes split and become easily differentiated even without signal processing. After performing the matched filtering process, which is explained in the Methods, we obtained the results of the normalized cross-correlation power between the measured waveforms in Fig 4c–d and the reference signal, which is the time-stretched waveform achieved in the single-target detection (see Fig. 3e). As depicted in

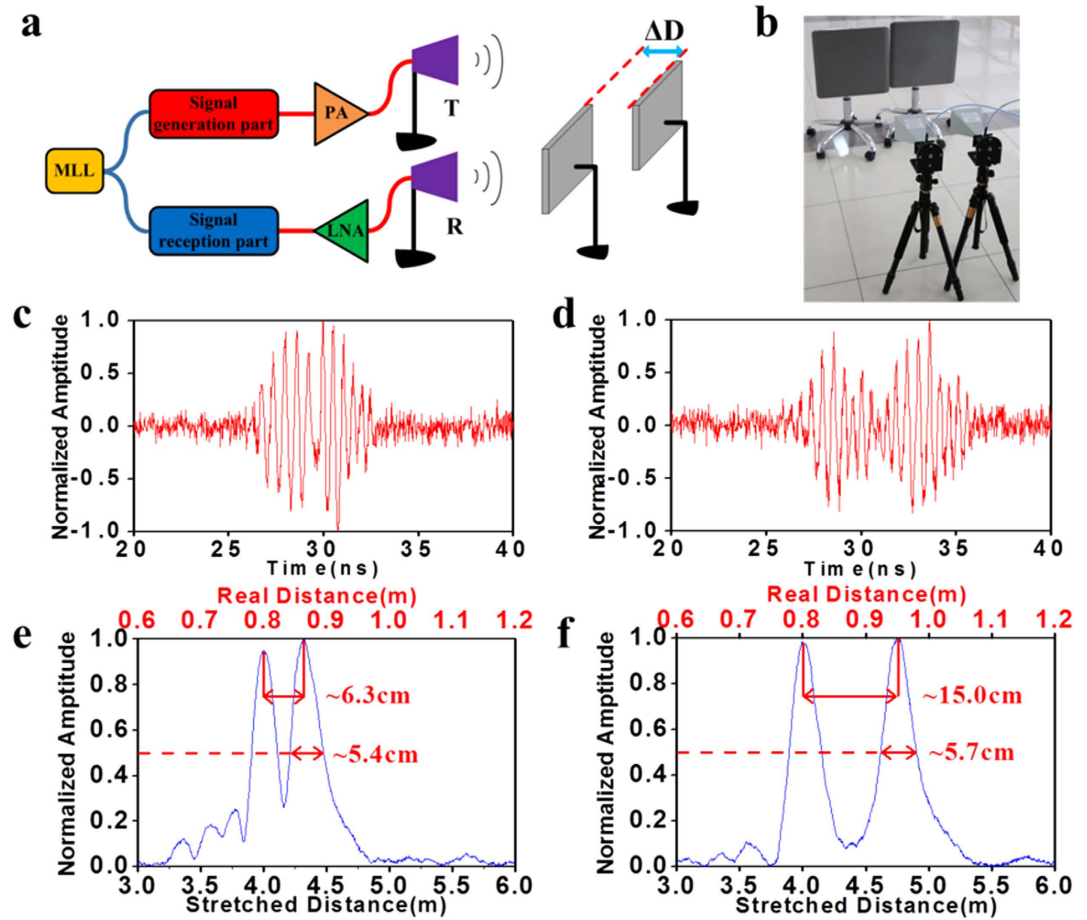

**Figure 4. Measurement results of the dual-target detection.** (**a**). Schematic diagram of the dual-target detection. (**b**). Experimental layout of the dual-target detection. (**c**). Received echo when the distance between the two targets is ~6.3 cm. (**d**). Received echo when the distance between the two targets is ~15.0 cm. (**e**). Result after the matched filtering process when the distance between the two targets is ~6.3 cm. (**f**). Result after the matched filtering process when the distance between the two targets is ~15.0 cm.

Fig. 4e–f, two reflection peaks are clearly distinguished in the ranging result. The coordinate in the bottom horizontal axis is 5-fold greater than that in the top horizontal axis; the coordinates represent the distance calculated using the time axis of Fig 4c–d and the real distance before the time stretch, respectively. The real distance corresponds to the radar's one-way transmitting distance from the targets to the transceiver or receiver antenna. The peak height is essentially influenced by the radar cross-section as well as the reflection angle of the metal plate. The left peak is positioned at ~80 cm, whereas the right peak moves after $\Delta D$ is increased. The measured distance between these two peaks indicates that the distance between the targets increases from ~6.3 cm to ~15.0 cm. The pulse widths at half maximum (i.e., ~5.4 cm and ~5.7 cm) are approximately equal to each other and represent the range resolution of the all-optical radar at the tailored X-band. This range resolution is maintained as the order of magnitude for the transmitted signal (centred at 10 GHz, 4 GHz bandwidth), whereas it is not determined by the time-stretched signal (centred at 2 GHz, 0.8 GHz bandwidth). Therefore, the detection ability of the system is not lost after the time-stretched reception.

## Discussion

The proposed all-optical central-frequency-programmable and bandwidth-tailorable radar architecture is strictly coherent because signal generation and reception originate from the same MLL as the optical source. Because of the seamless convergence of the wavelength-to-time mapping and time-stretch principle, high-resolution ranges at a low sampling rate are ensured. The radar architecture can be easily tailored for different central frequencies and different sweep bandwidths by simply adjusting the TOFs, and the electromagnetic properties of the targets do not suffer losses after the time-stretched processing. However, the utilization of the TOF1 and TOF2 in the generation component reduces the achievable pulse duration and time-bandwidth product since only a part of the chirped optical pulse is spectrally filtered with narrower bandwidth. If a dispersive loop is additionally introduced[29] and/or the TOF1 and TOF2 with much broader bandwidths have been replaced, both the pulse duration and time-bandwidth product might be further extended. With an appropriate PD, the frequency regions can be extended even higher to millimetre waves[22]. Although the tailored X-band is illustrated as an example for dual-target detection and detection ability verification, the effectiveness of a radar with a broader bandwidth and better range resolution at a higher central frequency is expected if the RF front-end (power amplifiers and antennas) is correspondingly substituted. Moreover, two antennas (transmission and reception) are used in this work, whereas a bistatic antenna, which was utilized in Ghelfi et al.[23], might also work properly. The measurement range of the radar (~4 m) is limited by the period of the MLL (27 ns) used in this work. However, the measurement range can be extended by using an external intensity modulator with a lower repetition rate laid after the MLL[29] or by substituting another MLL with a lower repetition rate. In the reception component, the continuous-time scheme of the photonic time-stretch ADC[27–28] may be adopted to ensure the successful modulation of the received echoes within the unambiguous range onto the pre-chirped optical pulses, even if prior information on the received echoes is not available. All of the above potential solutions to extend the time-bandwidth product in the generation component and adopt the continuous-time scheme of the photonic time-stretch ADC in the reception component are now under study.

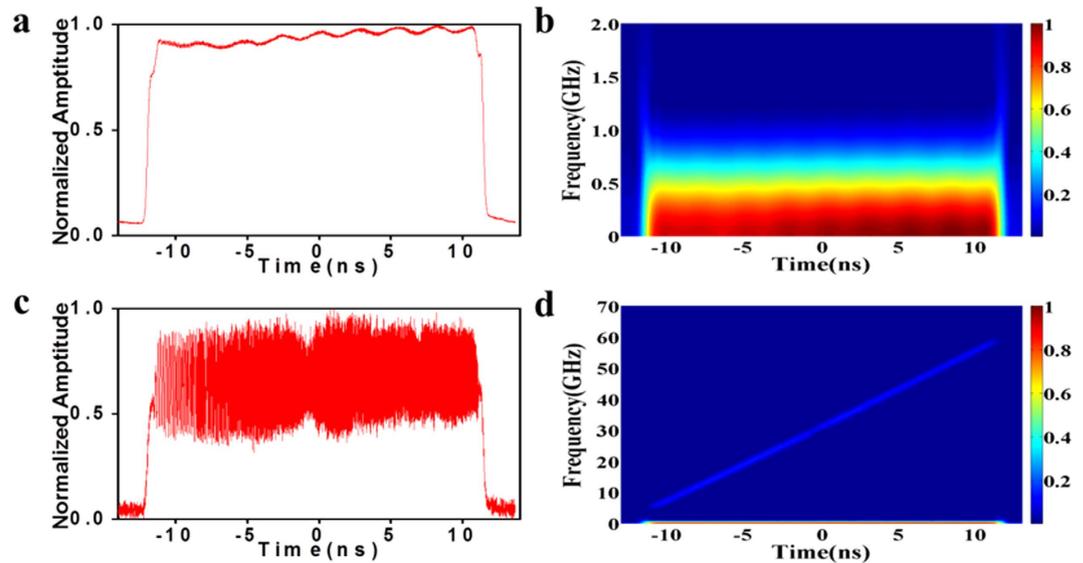

**Figure 5. Example of pulse-envelope elimination.** (**a**). Original waveform of the linearly chirped signal covering a wideband from 5 GHz to 60 GHz. (**b**). STFT analysis of the linearly chirped signal in (**a**,**c**). Average waveform of the linearly chirped signal. (**d**). Magnified view of the STFT analysis of the linearly chirped signal in the low frequency range.

## Methods

**Experimental components.** We used a mode-locked fibre laser (Precision Photonics Corp., FFL-1560-B) with a repetition rate of 37 MHz as the optical source. An oscilloscope (Agilent Technology Ltd., now Keysight Technology Ltd., DSAX96204Q) with a 63 GHz bandwidth was used to capture signals. Three tuneable optical filters (Alnair Labs, CVF-220CL) were used in the experimental setup, with two placed in the two arms to detect certain segments of the optical spectra for signal generation and the third used to generate the optical carrier for signal reception. In Fig. 2a–d, the dispersion value of DE1, filtering bandwidth of TOF1 and TOF2, and central wavelength difference between TOF1 and TOF2 are ~−3863 $ps^2$, ~8 nm, ~0.26 nm, respectively. A SMF of ~9 km was used as DE2. In Fig. 2e–g, the dispersion value of DE1 is ~−380 $ps^2$, and an ~200-meter SMF was used as DE2. The filtering bandwidths of TOF1 and TOF2 are ~2.4 nm (X-band), ~3.6 nm (Ku-band), and ~8.4 nm (Ka-band). The central wavelength differences between TOF1 and TOF2 are ~0.08 nm (X-band), ~0.12 nm (Ku-band), and ~0.26 nm (Ka-band). In Fig. 3, the horn antennas (LB-90-20-C-SF) for signal transmitting and receiving are mounted on two tripods. The power amplifier (Multilink Technology Corporation, MTC5515) and low noise amplifier (TLA-060120G36) both have a relatively flat gain at 8–12 GHz. The filtering bandwidth of TOF3 is 10 nm. The dispersion value of DE3 for the time-stretched processing is ~−1915 $ps^2$. PD1 (Finisar, XPDV4120R) and PD2 (Discovery DSC720) are used. The metal plates used as targets have a size of 40 cm × 40 cm.

**Pulse-envelope elimination.** Although the generated waveform is coherent, each pulse waveform is captured in real time by the oscilloscope. If the oscilloscope triggering is not exactly strict, the captured waveform is chaotic. Sufficient averaging of many pulse waveforms eliminates the radar signal of the instantaneous linear-chirped waveform and reproduces the stretched pulse envelope with the same repetition rate as the MLL. The stretched pulse envelope after thousands of times averaging has an approximately rectangular profile, and is illustrated in Fig. 5a, whereas its magnified STFT is depicted in Fig. 5b. Figure 5c illustrates one captured pulse waveform, including the generated radar signal of the instantaneous linearly chirped signal and the stretched pulse envelope. As shown in the short-time Fourier transform (STFT) analysis in Fig. 5d, strong low-frequency components occur because of the pulse envelope. Simple subtraction between Fig. 5c and Fig. 5a eliminates the pulse envelope. As illustrated in the temporal waveforms in Fig. 2, the linearly chirped waveform does not have a pulse envelope at the repetition rate of 37 MHz. Note that the pulse envelope with low-frequency components can also be equivalently eliminated by a high-bandwidth-pass electronic filter laid before or after the power amplifier and low noise amplifier. When processing the temporal waveforms in Fig. 3,4, we also use this numerical method to eliminate the pulse envelope.

**Matched filtering processing.** The matched filtering process of the received echoes in Fig. 4 is calculated based on the following formula:

$$y_{MF}(t) = x_{back}(t) \otimes x_{ref}^*(-t) \qquad (1)$$

where $x_{back}(t)$ represents the time-stretched echoes and $x_{ref}^*(-t)$ represents the conjugate of the time-reversed reference signal. The numerical implementation of the matched filtering process is described as follows:

(a) To perform the matched filtering processing, a reference waveform is needed. The reference waveform we used to calculate Fig. 4 is the waveform shown in Fig. 3e, which is the reflected echo acquired when a single target is placed in the beam path.
(b) The captured data using the oscilloscope only have a real value. The measured waveforms in Fig. 4c–d and the reference waveform are transformed to complex values through a Hilbert transformation to fulfil I/Q demodulation.
(c) The complex echoes are cross-correlated with the complex reference signal to fulfil the matched filtering according to Eq. (1). Consequently, the linearly chirped echoes are compressed to separate peaks.
(d) The pulse compression results are plotted against the time-stretched distance (bottom horizontal axis) and the real distance before the time-stretched processing (top horizontal axis).

### Acknowledgements
This work was supported in part by the National Natural Science Foundation of China (grant nos. 61571292, with 61535006, and 61127016), SRFDP of MOE (grant no. 20130073130005), and the State Key Lab Project of Shanghai Jiao Tong University under grant GKZD030033. The authors are grateful to Agilent Technologies Ltd. (now Keysight Technologies Ltd.) for lending us a high-speed real-time oscilloscope.

### Author Contributions
W.Z. and J.C. coordinated all the activities of the project. W.Z., H.Z. and J.C. designed the entire architecture, conducted the experiments, and wrote the paper. X.L. developed the data processing. S.Z. and Y.C. implemented the RF front and designed the targets.

### Additional Information
**Competing financial interests:** The authors declare no competing financial interests.

**How to cite this article**: Zou, W. *et al.* All-optical central-frequency-programmable and bandwidth-tailorable radar. *Sci. Rep.* **6**, 19786; doi: 10.1038/srep19786 (2016).